\documentclass{article}

\usepackage[dblblindworkshop, preprint]{neurips_2025}

\usepackage[utf8]{inputenc} 
\usepackage[T1]{fontenc}    
\usepackage{hyperref}       
\usepackage{url}            
\usepackage{booktabs}       
\usepackage{amsfonts}       
\usepackage{amsmath}        
\usepackage{nicefrac}       
\usepackage{microtype}      
\usepackage{xcolor}         
\usepackage{graphicx}
\usepackage{subcaption}

\workshoptitle{} 
\title{Quantum Noise Tomography with Physics-Informed Neural Networks}

\newcommand{\fig}[1]{Figure~\ref{#1}}

\newcommand{\savespace}{\vspace{-0.65em}}

\author{%
  Antonin Sulc\\
  Lawrence Berkeley National Lab, U.S.A.\\
  \texttt{asulc@lbl.gov} \\
}

\begin{document}
\maketitle

\begin{abstract}
Characterizing the environmental interactions of quantum systems is a critical bottleneck in the development of robust quantum technologies. Traditional tomographic methods are often data-intensive and struggle with scalability. In this work, we introduce a novel framework for performing Lindblad tomography using Physics-Informed Neural Networks (PINNs). By embedding the Lindblad master equation directly into the neural network's loss function, our approach simultaneously learns the quantum state's evolution and infers the underlying dissipation parameters from sparse, time-series measurement data. Our results show that PINNs can reconstruct both the system dynamics and the functional form of unknown noise parameters, presenting a sample-efficient and scalable solution for quantum device characterization. Ultimately, our method produces a fully-differentiable digital twin of a noisy quantum system by learning its governing master equation.
\end{abstract}

\savespace
\section{Introduction}
\savespace
Among the many challenges facing quantum computers, one of the most critical is decoherence~\cite{preskill2012quantum}, the process by which a quantum system gradually loses its quantum properties due to interactions with its surrounding environment.
Modeling open quantum system dynamics, governed by the Lindblad master equation~\cite{Lindblad1976}, is essential for designing high-fidelity quantum gates, developing error correction codes, and benchmarking quantum hardware. 
However, standard tomographic techniques for characterizing these dynamics, such as process tomography, require a number of measurements that scales exponentially with the number of qubits, rendering them intractable for even modest systems \cite{chuang1997prescription}. 

To address this challenge, we turn to the paradigm of Physics-Informed Neural Networks (PINNs) \cite{raissi2019physics}. 
PINNs are a class of neural networks that are trained to solve supervised learning tasks while respecting given laws of physics. By incorporating the physical model into the network's loss function, PINNs can effectively learn from sparse and noisy data, making them an ideal candidate for the data-scarce environment of quantum experiments. 

In this paper, we reframe Lindblad tomography from a black-box characterization problem to a physics-informed learning problem. 
Our key contribution is a framework that learns the governing differential equation for a system's dynamics from sparse data. This effectively creates an interpretable and predictive digital twin capable of not only reconstructing the state evolution but also discovering the functional form of unknown dissipative parameters.

\savespace
\section{Methodology}
\savespace
This section details our approach to quantum noise tomography using a PINN. The core of our method is the embedding of the Lindblad master equation directly into the network's loss function. This technique compels the model to find a physically valid solution that fits sparse data, enabling it to simultaneously reconstruct the system's dynamics and infer its unknown dissipation parameters 

\subsection{The Lindblad Master Equation}
\savespace
The evolution of an open quantum system's density matrix, $\rho$, is described by the Lindblad master equation~\cite{Lindblad1976}:
\begin{equation}
\frac{d\rho}{dt} = \underbrace{
-i[H, \rho] + \sum_k \gamma_k \left( L_k \rho L_k^\dagger - \frac{1}{2} \{L_k^\dagger L_k, \rho\} \right)
}_\text{Liouvillian super-operator $\mathcal{L}(\rho)$}
\label{eq:lindblad}
\end{equation}
\noindent where $H$ is the system Hamiltonian governing the coherent evolution, and the second term is the dissipator $\mathcal{D}(\rho)$ which describes the incoherent, environmental effects. The $L_k$ are the Lindblad or "jump" operators, and the $\gamma_k \ge 0$ are the corresponding dissipation rates. The central challenge of Lindblad tomography is to determine the set of operators $\{L_k\}$ and rates $\{\gamma_k\}$ from experimental data.

\savespace
\subsection{Sources of Decoherence}
\savespace
We model two prevalent noise channels. We model amplitude damping ($\gamma_{AD}$), which models energy relaxation, with jump operator $\hat{L}_1 = \hat{\sigma}^- = |0\rangle\langle 1|$. Furthermore, we also model pure dephasing ($\gamma_{PD}$), which models loss of phase coherence, with jump operator $\hat{L}_2 = \hat{\sigma}_z = \text{diag}(1, -1)$. 

\savespace
\subsection{Physics-Informed Neural Networks}
\savespace
Our approach models the system's state using a neural network $f$ with parameters $\theta$. The network takes time $t$ as input and outputs the expectation values of a complete basis of operators (e.g., Pauli operators), from which the density matrix $\rho_\theta(t)$ can be reconstructed. The dissipation rates, $\{\gamma_k\}$, are treated as learnable parameters. the neural network is trained to predict the system's key observables over time, from which the full density matrix $\rho_\theta(t)$ is then constructed.

The network is trained by minimizing a total loss function, $w_d L_{data} + w_p L_{phys} + w_{ic} L_{ic}.$, which consists of three main components:

\savespace
\paragraph{Data Loss ($L_d$)} A standard mean squared error loss that penalizes deviations between the network's predictions and the sparse experimental data points.
    \begin{equation}
    L_{data} = \frac{1}{N_d} \sum_{i=1}^{N_d} \left\| \text{Tr}(\sigma_j \rho_\theta(t_i)) - \langle\sigma_j\rangle_{t_i}^{\text{data}} \right\| ^2
    \end{equation}

\savespace
\paragraph{Physics Loss ($L_{phys}$):} The core of the PINN. This loss enforces the Lindblad master equation. 
We use automatic differentiation to compute the time-derivative of the density matrix $\frac{d\rho_\theta}{dt}$, and construct the dissipator using the learnable rates $\gamma_k$.
The loss is the mean squared norm of the master equation's residual, evaluated at a set of collocation points distributed throughout the time domain.
    \begin{equation}
    L_{phys} = \frac{1}{N_c} \sum_{i=1}^{N_c}\left\| \frac{d\rho_\theta(t_i)}{dt} - \mathcal{L}(\rho_\theta(t_i)) \right\|_F^2
    \end{equation}
We stress that $L_{phys}$ does not rely on the input data points. It imposes the physical consistency given by Equation~\ref{eq:lindblad}. Generating these points is very cheap as the only constraint is that they must be in the time interval $t$.

\savespace
\paragraph{Initial Condition Loss ($L_{ic}$):} Ensures the network's solution satisfies the known initial state of the system, $\rho(0)$.

By minimizing this total loss, the network learns a physically valid trajectory $\rho_\theta(t)$ that fits the data, and the parameters $\gamma_k$ converge to their true values. We also include a penalty term to enforce the positivity of the density matrix.

\savespace
\section{Experiments, Simulations and Results}
\savespace
We present results from simulated experiments designed to test our framework's accuracy and data efficiency. For each case, the training data consists of sparse measurements of Pauli operator expectation values over a time interval of $t \in [0, 5.0]$. To assess the model's robustness against realistic experimental conditions, we evaluate its performance on both ideal, noise-free data and data corrupted by additive Gaussian noise, defined as $\langle\sigma_j\rangle_{t_i}^{\text{noisy}} = \langle\sigma_j\rangle_{t_i}^{\text{true}} + \epsilon$, where $\epsilon \sim \mathcal{N}(0, \sigma^2)$.

This section demonstrates the framework's effectiveness across three simulated scenarios of increasing complexity: a single qubit with time-varying noise, a two-qubit system with local noise under various measurement error levels, and a highly entangled two-qubit system.

\savespace
\subsection{Case 1: Single Qubit, Time-Varying Rates}
\begin{figure}[h!]
  \centering
  \resizebox{1.0\linewidth}{!}{
  \begin{tabular}{cc}
  \includegraphics[width=1.0\linewidth]{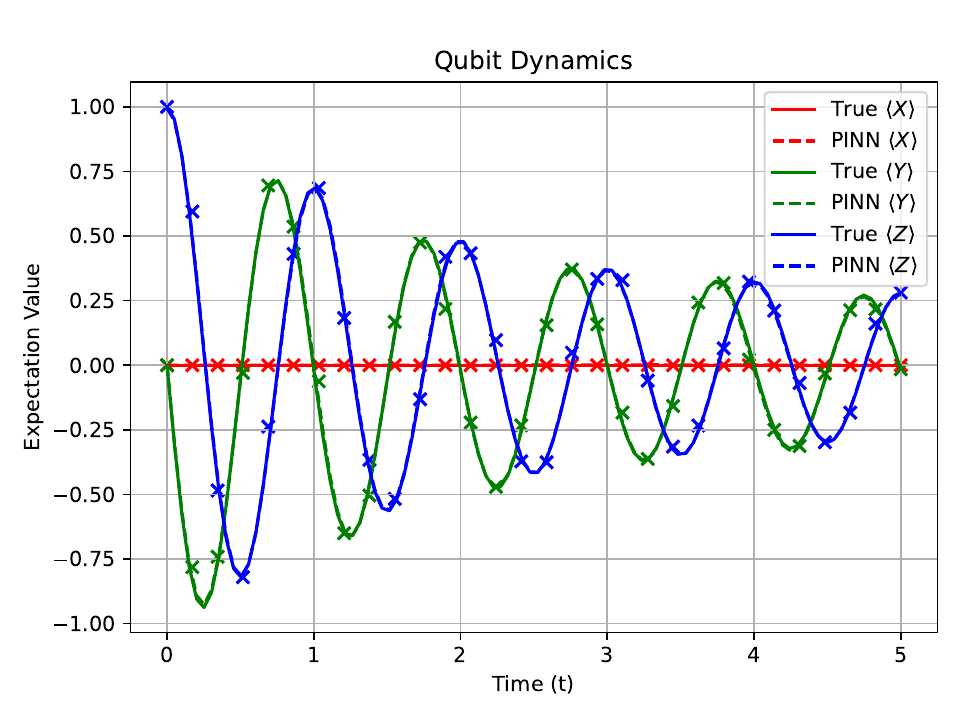} &\includegraphics[width=1.0\linewidth]{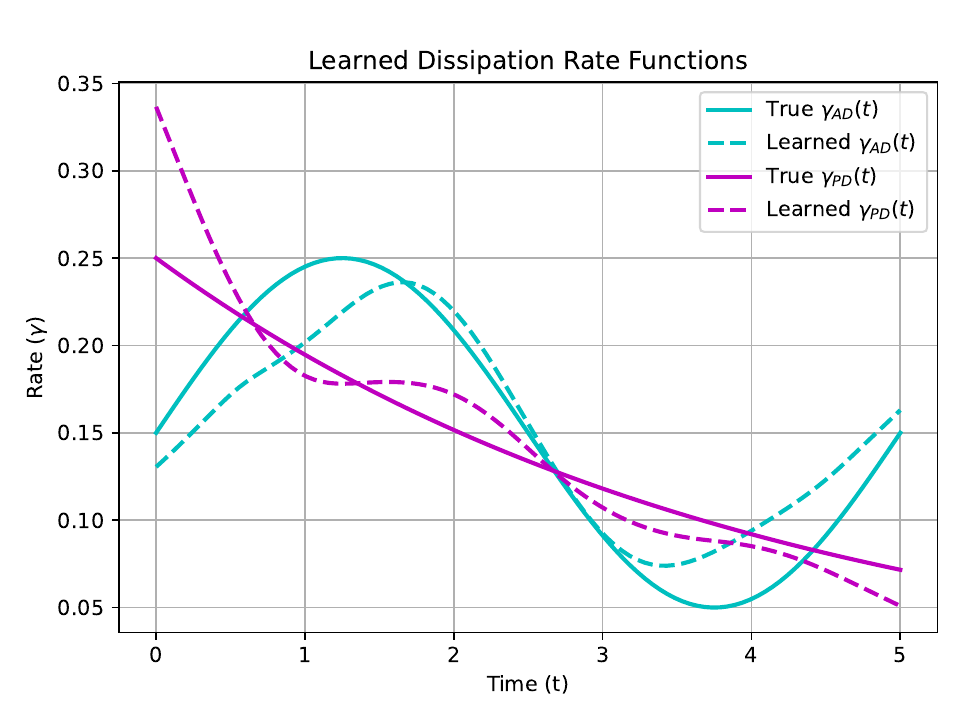}
  \end{tabular}}
  \caption{\textbf{Single Qubit with Time-Varying Dissipation.} (Left) The PINN learns the system dynamics. Notice randomly and sparsely distributed training points distinguished with $\times$ symbols. (Right) The network discovers the unknown functional forms of the time-dependent dissipation rates $\gamma_{AD}(t)$ and $\gamma_{PD}(t)$.}
  \label{fig:1q_time}
\end{figure}

\savespace
This example involves noise processes that vary over time and is solved using a residual neural network trained with the Adam optimizer. It is trained on 30 data points of Pauli observables $\langle X \rangle, \langle Y \rangle, \langle Z \rangle$. 
The PINN architecture is tailored to output time-dependent functions $\gamma_{AD}(t)$ and $\gamma_{PD}(t)$. \fig{fig:1q_time} shows that from 30 sparse data points alone, the PINN reconstructs the underlying functional forms of the dissipation rates—a sinusoidal and an exponential decay—without any prior knowledge.

\savespace
\subsection{Case 2: Two-Qubit System}
\begin{figure}[h!]
  \centering
  \resizebox{1.0\linewidth}{!}{
  \begin{tabular}{cccc}
  {\small Noise $\sigma = 0.05$ - $\langle X_1 \rangle, \langle Z_1 \rangle$} & {\small Noise $\sigma = 0.05$ - $\langle X_2 \rangle, \langle Z_2 \rangle$} & {\small Noise $\sigma = 0.1$ - $\langle X_1 \rangle, \langle Z_1 \rangle$} & {\small Noise $\sigma = 0.1$ - $\langle X_2 \rangle, \langle Z_2 \rangle$}\\
  \includegraphics[width=0.25\linewidth]{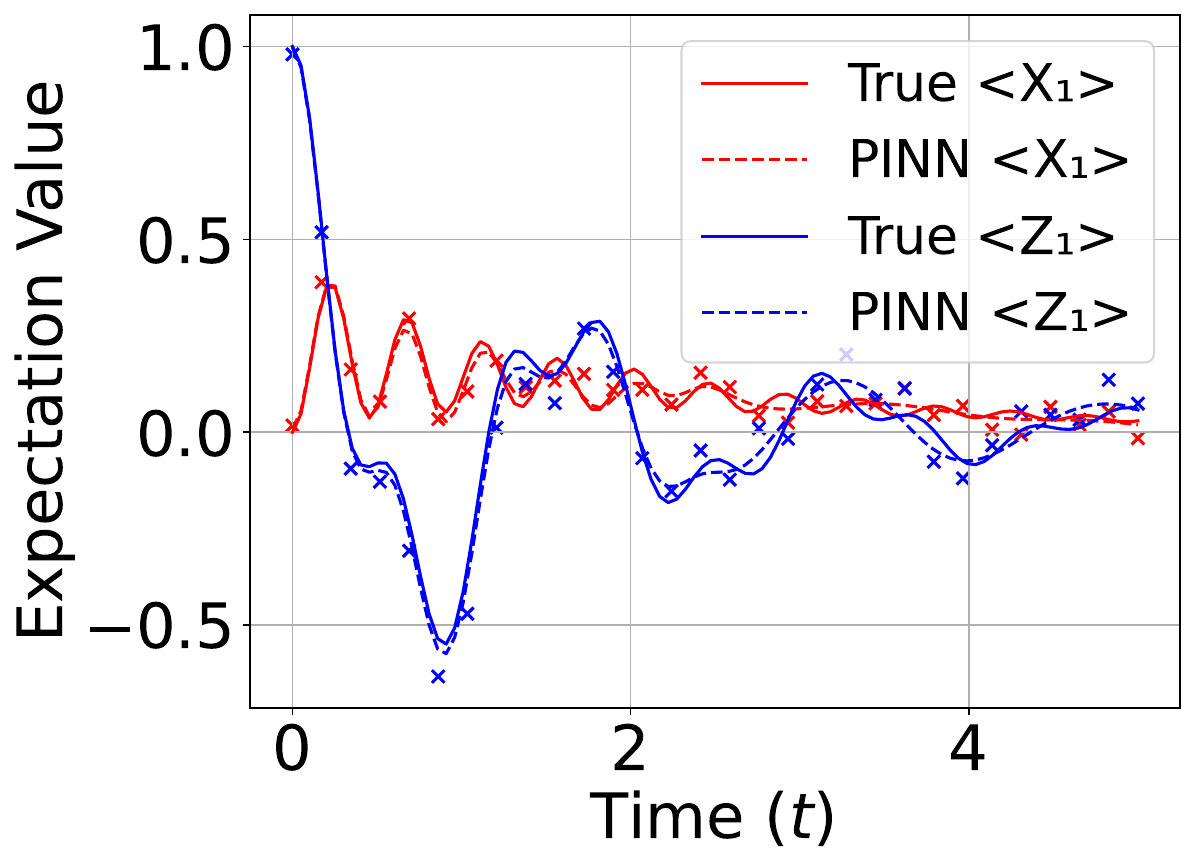} &
  \includegraphics[width=0.25\linewidth]{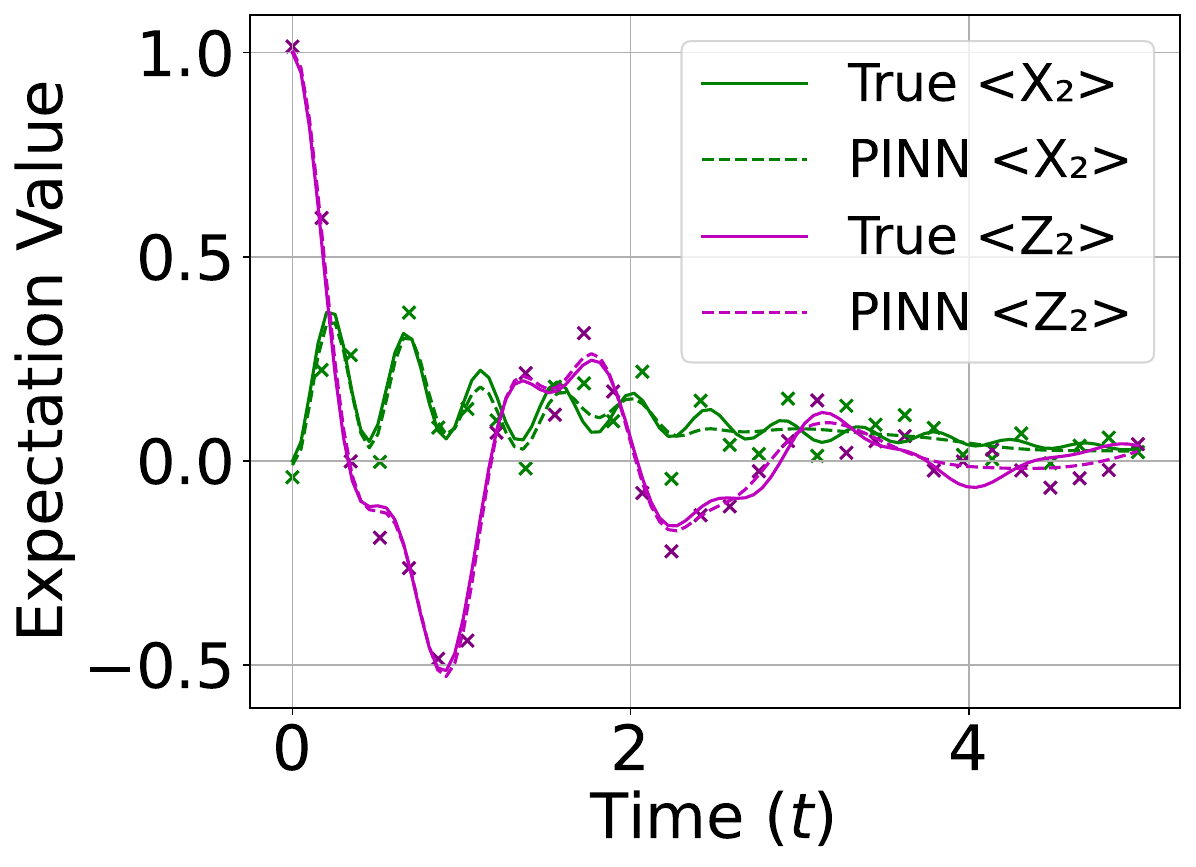} &
  \includegraphics[width=0.25\linewidth]{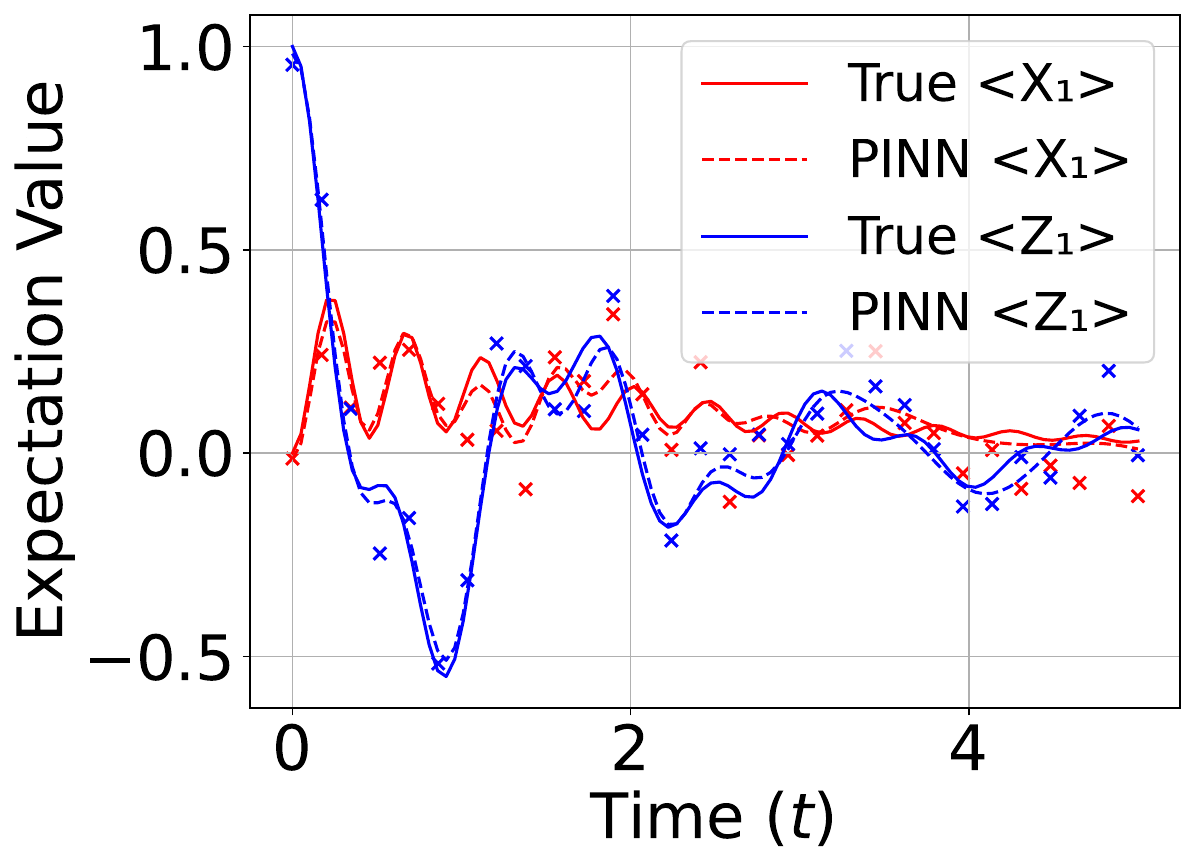} &
  \includegraphics[width=0.25\linewidth]{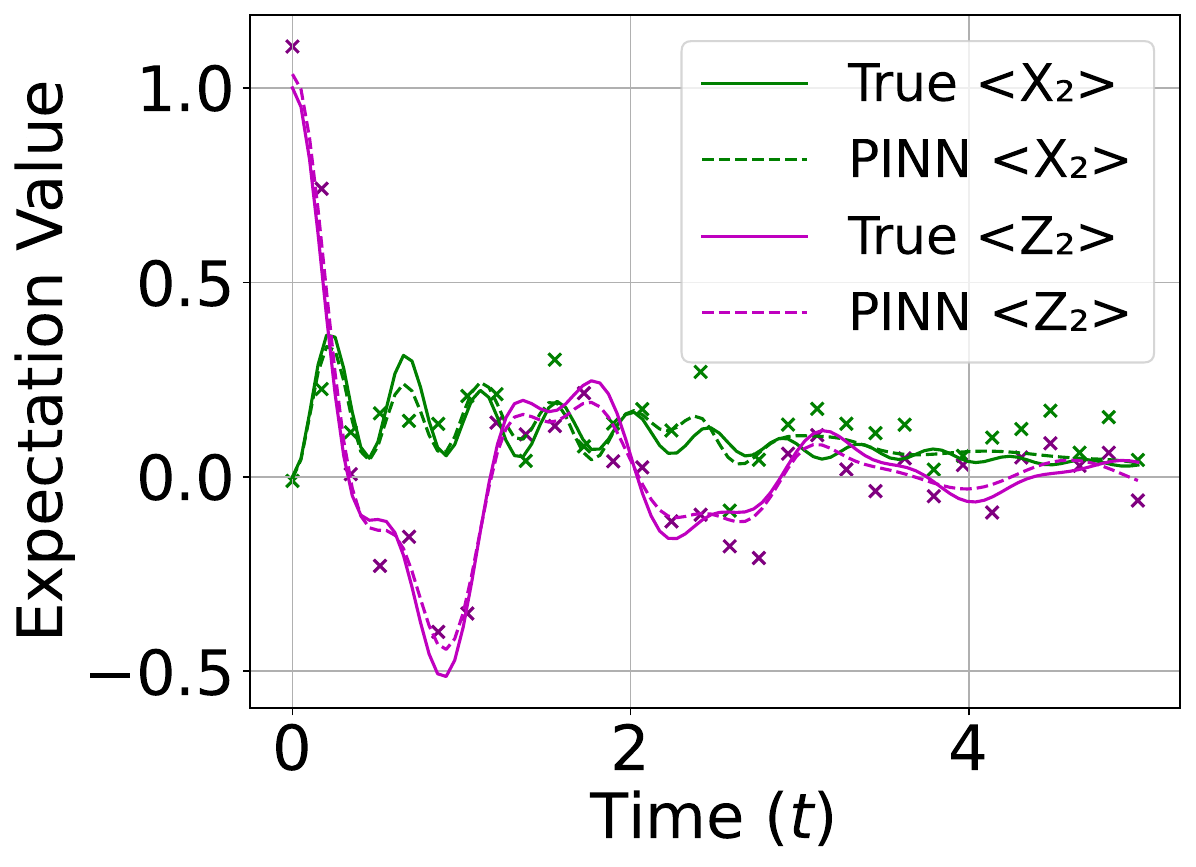}
  \end{tabular}}
  \caption{Two-Qubit System with Local Noise: The PINN correctly learns the local qubit dynamics from sparse data for both the low noise case $\mathcal{N}(0,0.05)$ (left two panels) and when the training data is corrupted with Gaussian noise $\mathcal{N}(0,0.1)$(right two panels).}
  \label{fig:2q_local}
\end{figure}

\savespace
We then scaled our method to a two-qubit system with four independent, local noise channels, training on only 25 data points of local observables ($\langle X_1 \rangle, \langle Z_1 \rangle, \langle X_2 \rangle, \langle Z_2 \rangle$). The PINN infers the four constant dissipation rates, $\gamma_k$. In addition to the noise-free case, we tested the model's robustness by adding Gaussian noise with a standard deviation $\sigma$ of 0.01, 0.025, 0.05, 0.1, and 0.2 to the sparse training points. As shown in Figure~\ref{fig:2q_local}, the PINN accurately reproduces the system dynamics even in the presence of measurement noise on training data (e.g., for $\sigma=0.1$). Table~\ref{tab:stds} provides a quantitative analysis of the parameter estimation, demonstrating that the inferred dissipation rates remain close to their true values for low noise levels, with performance degrading as noise increases.

\savespace
\subsection{Case 3: Two-Qubit Entangling System}
\savespace
Finally, we tested the PINN on a two-qubit system with a Hamiltonian designed to generate high entanglement, a critical resource for quantum algorithms. 
For this case, since the highly entangled-state dynamics have many high-frequency components, the architecture uses Fourier features~\cite{tancik2020fourier} before three layers of residual blocks and was trained with AdamW and refined by L-BFGS.

\begin{figure}[h!]
\centering
\begin{minipage}[t]{0.5\linewidth}
    \centering
    \resizebox{\linewidth}{!}{
    \begin{tabular}{cc}
        $\langle X_1 \rangle, \langle Z_1 \rangle$ & $\langle X_2 \rangle, \langle Z_2 \rangle$\\
        \includegraphics[width=1.0\linewidth]{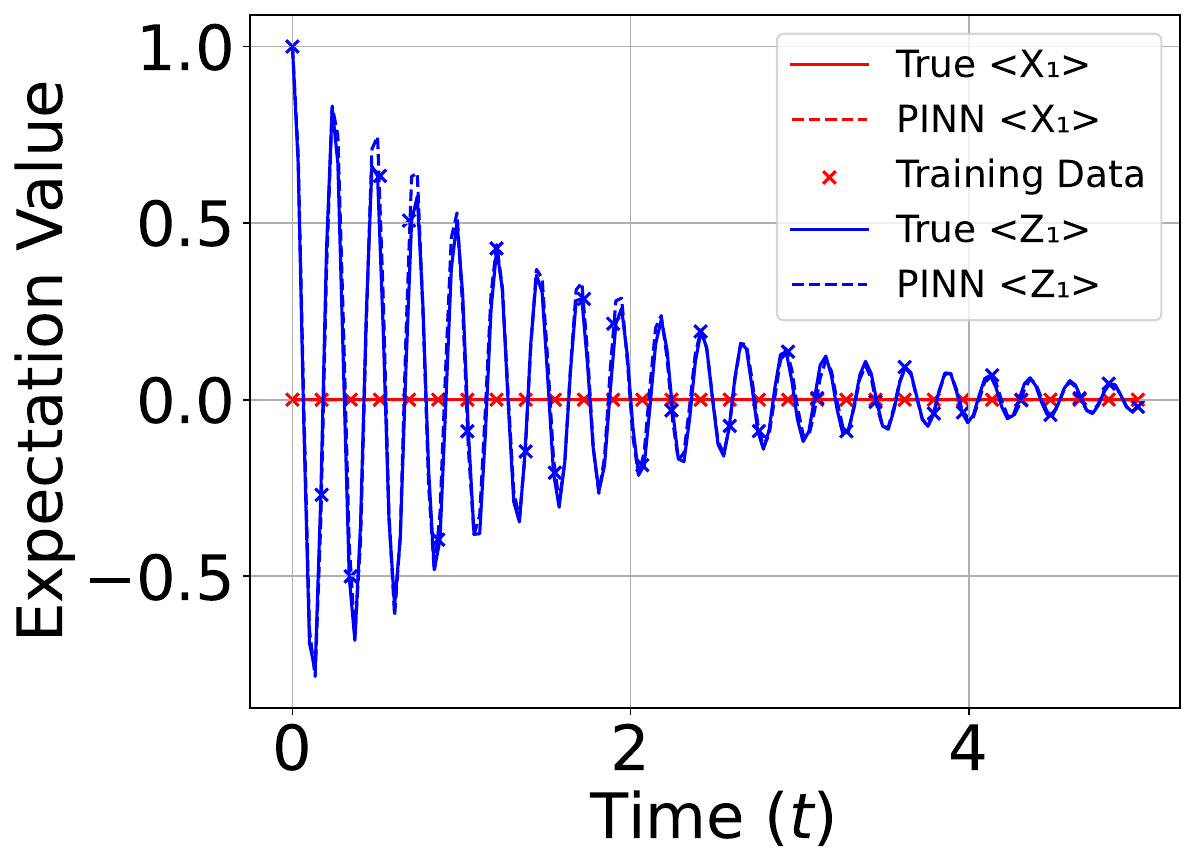} &
        \includegraphics[width=1.0\linewidth]{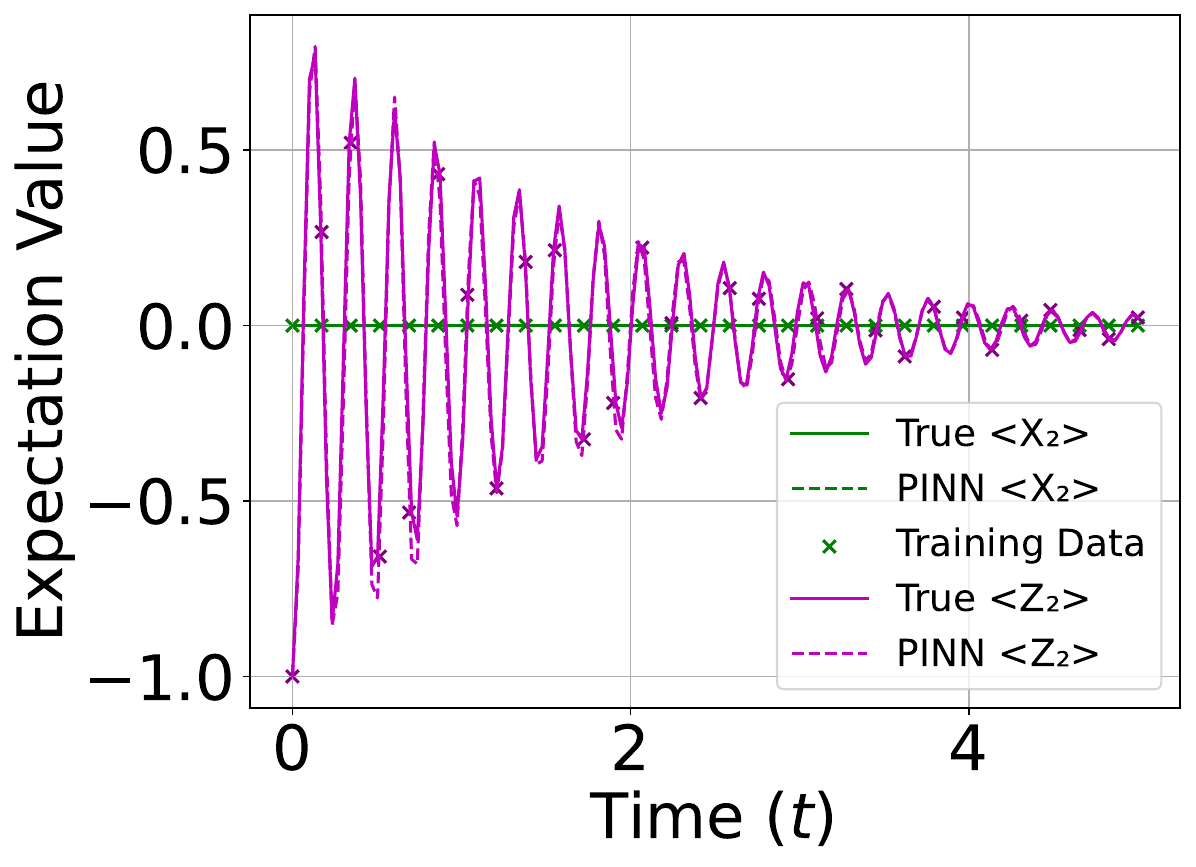}
    \end{tabular}
    }
    \caption{Two-Qubit High-Entanglement System: The PINN correctly learns the dynamics of local observables ($\langle X_1 \rangle, \langle Z_1 \rangle$ and $\langle X_2 \rangle, \langle Z_2 \rangle$) for a system designed to generate strong entanglement. 
    The model is trained on 30 sparse data points (marked with $\times$).}
    \label{fig:2q_entangle}
\end{minipage}%
\hfill
\begin{minipage}[t]{0.45\linewidth}
    \centering
    \resizebox{1.0\linewidth}{!}{
    \begin{tabular}{l|cccc}
        \toprule
         & $\gamma_{AD1}$ & $\gamma_{PD1}$ & $\gamma_{AD2}$ & $\gamma_{PD2}$\\ 
         \midrule
         True rates & 0.2500 & 0.1500 & 0.1000 & 0.3000\\
         \hline
         No noise & 0.2857 & 0.1399 & 0.1341 & 0.2893\\
         $\sigma =0.01$ & 0.2209 & 0.2135 & 0.1028 & 0.3259\\
         $\sigma=0.025$ & 0.2232 & 0.1824 & 0.1555 & 0.2831\\
         $\sigma=0.05$ & 0.2117 & 0.1823 & 0.1793 & 0.2535\\
         $\sigma=0.1$ & 0.1884 & 0.1586 & 0.0545 & 0.4537\\
         $\sigma=0.2$ & 0.0000 & 0.1853 & 0.7350 & 0.0000\\
         \hline
         Entangled System & 0.3624 & 0.1061 & 0.0022 & 0.2562 \\
         \bottomrule
    \end{tabular}}
    \captionof{table}{Inferred Dissipation Rates for Two-Qubit Systems: The table compares the true dissipation rates with the values inferred by the PINN. The results correspond to the local noise case (Case 2) under varying levels of additive Gaussian noise (standard deviation $\sigma$) and the high-entanglement system (Case 3, final row). The model shows performance degrading as noise increases for $\sigma>0.05$}
    \label{tab:stds}
\end{minipage}
\end{figure}

The model is trained on 30 data points of local observables $\langle X_1 \rangle, \langle Z_1 \rangle, \langle X_2 \rangle, \langle Z_2 \rangle$.
Characterizing noise in the presence of strong correlations is very difficult. As seen in Table~\ref{tab:stds} the PINN remains accurate on most variables ($\gamma_{AD1}, \gamma_{PD1},\gamma_{PD2}$), it struggles to learn $\gamma_{AD2}$ in this highly correlated regime while correctly capturing the complex, non-local dynamics and the generation of entanglement (concurrence) as \fig{fig:2q_entangle} shows.

\savespace
\section{Novelty, Future Work, Limitations and Discussion}
\savespace
The novelty of this work lies in framing Lindblad tomography as a physics-informed inverse problem. 
Our PINN-based approach is data-efficient, requiring only very few points to learn accurate models from sparse data, thereby reducing the experimental overhead of device characterization. 

By avoiding the exponential measurement requirements of standard tomography, this approach presents a promising path towards a characterization tool that scales polynomially with the number of qubits. Additionally, the trained PINN acts as a fully-differentiable digital twin of the quantum device, enabling applications such as simulation, optimal control design, and the prototyping of tailored error mitigation strategies.
Such a digital twin provides a tool for mitigating noise-induced barren plateaus~\cite{larocca2025barren}, which is a primary bottleneck where hardware noise exponentially flattens the training landscape, as the precise noise model can inform the design of resilient algorithms.
However, our experiments are still very limited to small setups restricted to simulations, and despite testing robustness of the model against measurement noise, real hardware poses many underlying challenges. 

As mentioned, a trained PINN provides a differentiable model of the noise, which can be used to correct experimental data via advanced techniques. Because the PINN learns the specific noise of the device, it can enable tailored, high-efficacy QEM strategies that outperform generic approaches.

\savespace
\section{Conclusion}
\savespace
We have presented a novel and powerful framework for characterizing open quantum systems using Physics-Informed Neural Networks. By leveraging the structure of the Lindblad master equation as a physical constraint, our method can learn both the system dynamics and the underlying dissipation parameters from sparse data. 
Our simulated experiments demonstrate the efficacy of this method across single and two-qubit systems, including the challenging case of time-varying noise. Crucially, we also show that our PINN-based approach is robust to simulated measurement noise, accurately inferring dissipation parameters even when trained on corrupted data. This data-efficient and scalable framework represents a promising modeling tool for quantum tomography. Future work will focus on deploying this method on real quantum hardware and exploring its application in designing tailored error mitigation protocols.

\clearpage
\bibliographystyle{unsrt}
\bibliography{references}

\clearpage
\appendix
\section{Model Architecture and Training Details}
\subsection{Network Architecture}
For all experiments, we used a multi-layer perceptron (MLP) with residual connections. The base architecture consists of an input layer, a series of residual blocks, and an output layer. Each residual block contains two linear layers separated by a SiLU activation function.
\begin{itemize}
    \item \textbf{Case 1 \& 2:} The network used 2 residual blocks with 64 hidden features.
    \item \textbf{Case 3:} To capture the high-frequency dynamics, the input time variable was first transformed using a Fourier feature mapping layer~\cite{tancik2020fourier}~before being passed to a deeper network of 3 residual blocks with 64 hidden features.
\end{itemize}

\subsection{Training and Optimization}
The networks were trained for 100,000 epochs.
\begin{itemize}
    \item \textbf{Case 1 \& 2:} We used the Adam optimizer with a learning rate of $1 \times 10^{-3}$.
    \item \textbf{Case 3:} We used the AdamW optimizer with a learning rate of $1 \times 10^{-3}$, followed by a refinement step using the L-BFGS optimizer to achieve final convergence. An exponential learning rate scheduler with $\gamma=0.999$ was used during AdamW training.
\end{itemize}
The weights for the composite loss function were set to $w_d=1.0, w_p=10.0, w_{ic}=1.0$ for all experiments. A positivity constraint on the density matrix was enforced by adding a penalty term, $L_{constr}$, to the total loss. This term is calculated by taking the mean of the rectified negative eigenvalues of the predicted density matrix $\rho_\theta(t)$ at each collocation point:
\begin{equation}
    L_{constr} = \frac{1}{N_c N_{dim}} \sum_{i=1}^{N_c} \sum_{j=1}^{N_{dim}} \text{ReLU}(-\lambda_j(\rho_\theta(t_i)))
\end{equation}
where $\lambda_j(\rho)$ are the eigenvalues of the density matrix $\rho$, $N_c$ is the number of collocation points, and $N_{dim}$ is the dimension of the Hilbert space. This ensures that the network is penalized for producing physically invalid, non-positive semi-definite states.

The training was performed on Google Colab Environment on Intel Xeon 2.20GHz CPU equpped with 80 GB RAM and A100 40 GB NVIDIA GPU. 

\end{document}